\documentstyle[12pt]{article}                                          

\begin{document}
                                                                       
\begin{titlepage}
\begin{center}                                                         
{\LARGE\bf Statistical Models on Spherical Geometries}

\vspace{2.0cm}                                                         
\large{
S. Boettcher$^*$ and M. Moshe\\
Physics Department\\
Technion -- Israel Institute of Technology, Haifa 32000, ISRAEL\\ }

\end{center}                                                           
\vspace{4.0cm}                                                         
\abstract{
We use a one-dimensional random walk on $D$-dimensional hyper-spheres
to determine the critical behavior of statistical systems in 
hyper-spherical geometries. First, we demonstrate the properties of 
such a walk by studying the phase diagram of a percolation problem. We find a
line of second and first order phase transitions separated by a tricritical
point. Then, we analyze the adsorption-desorption transition for a polymer 
growing near the attractive boundary of a cylindrical cell membrane. We find 
that the fraction of adsorbed monomers on the boundary vanishes exponentially 
when the adsorption energy decreases towards its critical value. 
}
\vfill
\noindent
PACS numbers: 05.20.-y, 05.40.+j, 05.50.+q

\vspace{2.0cm}                                                         
\noindent
{$^*$} {\it Permanent Address: Department of Physics, 
Brookhaven National Laboratory, Upton, NY 11973, USA.} 
\end{titlepage}

The study of random walk models in statistical mechanics has led to a
variety of practical applications\cite{AIP} and illuminating analytical 
results.\cite{Priv1} They provide many insights into the nature of 
critical phenomena observed in more complicated systems. For
instance, in statistical field theory, the success of random walk models 
is best epitomized by its application to the triviality problem.
\cite{Aizenman,Sokal} A fruitful area of application closer to the 
subject of this letter concerns the study of polymers using self-avoiding
walks.\cite{FloMo}

In this letter we propose a new model of random walks that substantially 
simplifies the study of statistical systems in arbitrary dimensions. 
In general, in systems with spherical symmetry we can decouple radial 
and angular motion. For such systems it is often found that many of the 
qualitative features of the dynamics are determined by the radial component 
alone. We show that nontrivial critical properties of physical quantities 
depending only on a radial separation are obtained with our model of
random walks on concentric hyper-spheres. 

With this model we also extend the powerful techniques developed for random 
walk models to access the plethora of phenomena that spatial boundary 
conditions induce.\cite{Binder} It is well known that the introduction of a 
boundary can alter the physical character of a statistical 
system.\cite{Dimarzio1} As an example of the subtle
effects that our method reveals for statistical systems near curved
boundaries, we discuss the adsorption-desorption transition of polymers
growing near an attractive boundary.\cite{Taka}
In the limit of an infinitely extended polymer, a finite fraction
$P(\kappa)$ of monomers gets adsorbed on the boundary as soon as the
attractive potential $\kappa$ on the boundary increases above a critical
value. For planar boundaries one generically finds that $P(\kappa)$ vanishes
linearly when $\kappa$ approaches $\kappa_c$ in the adsorbed phase. With 
our model we can study such a system near a curved boundary like a cylindrical
cell membrane.\cite{Dimarzio2} Here, the configurational entropy due to the ``open space''
available to the polymer chain is bigger than that in the neighborhood of a
flat boundary. Accordingly, one would expect that the transition is
weaker. In fact, with our model we show that for a cylinder of radius 
$m\geq 0$ in monomer units and for 
$\Delta \kappa\equiv \kappa-\kappa_c \to 0_+$, the asymptotic behavior
of $P(\kappa)$ is given by
\begin{equation}
P(\kappa)\sim {4\over 81} 
{e^{-{8\over 9(m+1) \Delta \kappa}}\over (m+1)\Delta \kappa^2}.
\end{equation}
This asymptotic expression ceases to be valid when 
$\Delta \kappa\sim 0.188/(m+1)$ where we observe a crossover to linear 
growth in the exact expression of $P(\kappa)$.

In this letter we first describe our model of random walks on a 
hyper-spherical geometry. We calculate the phase diagram for a simple 
percolation model to demonstrate the effects that random walks
in such a geometry bring about. Then, we present our calculation 
of the adsorption transition 
for the case of a cylindrical geometry which leads to Eq.~(1). 

Consider an infinite set of concentric and equally spaced spheres in 
arbitrary spatial dimension $D$. $S_n$, $n=0,~1,~2,~3,\ldots,$ designates the 
surface of the $n$th sphere from the center with area $2\pi^{D/2}n^{D-
1}/\Gamma(D/2)$. We define a random walk on such a configuration in the
following way:\cite{BeBoMe} Let $c_{n,t}$, $t\geq0,~n\geq1,$ be the
probability for a random walker to be located {\sl anywhere} between
$S_{n-1}$ and $S_n$ at time step $t$. In this way, the angular
position of the walker with respect to the origin is completely averaged
away. The walker may have the choice to
stay inside the region between $S_{n-1}$ and $S_n$ with probability
$P_{stay}(n)$, walk outward with a probability proportional to the total
surface area of the sphere just outward, $P_{out}(n)= n^{D-1}/{\cal N}(n)$, 
and similarly inward with $P_{in}(n)=(n-1)^{D-1}/{\cal N}(n)$, 
$n\geq 2$, where the norm ${\cal N}$ is defined through $P_{out}
+P_{in}+P_{stay}\equiv 1$. Thus, we find a $1+1$-dimensional evolution
equation to describe the behavior of the walker in this $D$-dimensional
geometry:
\begin{equation}
c_{n,t}=P_{out}(n-1)~c_{n-1,t-1} + P_{stay}(n)~c_{n,t-1} + P_{in}
(n+1)~c_{n+1,t-1}.
\end{equation}
To be consistent with Ref.~10, we designate $n=1$ to be the innermost 
region, leading to the boundary condition $P_{in}(1)=0$ (which is implicit 
for $D>1$). In Ref.~10, the properties of Eq.~(2) were analyzed for the
initial condition $c_{n,0}=\delta_{n,0}$ and arbitrary $P_{out}$ and $P_{in}$ 
with $P_{stay}\equiv 0$. For example, for the case of concentric spheres it
was found that the probability of ever returning to the origin for a
random walker starting at the origin is given by $\Pi_D = 1-1/\zeta(D-
1)$ for $D>2$,\footnote{$\zeta$ refers to Riemann's $\zeta$ function.} 
and unity for $D\leq 2$. This result is in qualitative agreement with
$\Pi_D$ for a hyper-cubic lattice.\cite{ItDr} In fact, it has been
shown that one-dimensional random walks on a hyper-spherical lattice 
have the same scaling behavior as $D$-dimensional random walks on a 
hyper-cubic lattice.\cite{BeBoMo}

These ideas can be extend to the following
directed percolation problem: For $t=0$, there is one ``wet''
site on an infinitely extended line of ``dry'' sites with unit spacing.
At even times $t$, sites take on only integer values, at odd times $t$,
sites take on only half-integer values, $\pm 1/2,~\pm 3/2,\ldots$. If
at time $t-1$ two neighboring sites at $i$ and $i+1$ are ``dry'', then
at time $t$ the site at $i+1/2$ is also ``dry''. Furthermore, we assume
that two ``wet'' sites at $i$ and $i+1$ at time $t-1$ {\it always} produce a
``wet'' site at $i+1/2$ at time $t$. The latter assumption makes the
percolation cluster, i. e. the region of all ``wet'' sites at any time
$t$, compact and there are only two interfaces between the compact
``wet'' cluster in the middle and the two surrounding ``dry'' regions.
Thus, we merely have to specify the following three situations: On the
next time step the gap between the interfaces either makes a unit step
outward with a probability $P_{out}$, or it makes a unit step inward
with a probability $P_{in}$, or both interfaces shift to the right or
to the left without widening the gap with a total probability $P_{stay}$.
The behavior of such a compact percolation cluster can be mapped 
into the one-dimensional random walk in Eq.~(2) by defining $c_{n,t}$ to 
be the probability that the gap between both interfaces at time $t\geq 0$ 
is of width $n\geq 0$.\cite{DoKi} Note that $c_{n,t}$ is now defined for all 
$n\geq 0$, implying the boundary condition $P_{in}(0)=0$ here. The percolation
probability $P$ is then defined as the probability of the cluster to persist 
for all times $t$, i. e. $n>0$ for all $t$. Thus, $P$ is the complement of 
the probability of ever returning to the origin for a random walker in 
Eq.~(2). Essam\cite{Essam} has analyzed this problem in depth for the case 
$P_{out}=q^2$, $P_{in}=(1-q)^2$, and $P_{stay}=2q(1-q)$ where $q$, $0\leq q
\leq 1,$ is a constant independent of the width of the cluster. Generally,
we find the percolation probability [see also Eq.~(3.16) in Ref.~10]
\begin{equation}
P= \left[\sum_{n=1}^{\infty} \prod_{i=1}^{n-1} {P_{in}(i) \over
P_{out}(i)}\right]^{-1},
\end{equation}
reducing to Essam's result: $P(q)=(2q-1)/q^2$ for $q>1/2$, and $P(q)\equiv 0$ 
for $q \leq 1/2$. 

More interesting results can be generated from Eq.~(3) by choosing 
distance-dependent coefficient functions $P_{out}$ and $P_{in}$. [Note that
Eq.~(3) does not depend on $P_{stay}$.] For instance, for
$P_{out}(n)=q^2 (n+1)^{\delta}/{\cal N}(n)$ and
$P_{in}(n)=(1-q)^2 n^{\delta}/{\cal N}(n)$, $n\geq 1$, with $P_{out}
(0)=1$, we obtain $\delta$-dependent critical coefficients. We compute the 
percolation probability\cite{Bateman}
\begin{equation}
P(q,\delta)=\left[\sum_{n=0}^{\infty} 
\left ({\scriptstyle{1-q \over q}} \right)^{2n}\,
(n+1)^{-\delta}\right]^{-1}
=\Phi\left[\left({\scriptstyle{1-q \over q}}\right)^2,\delta,1\right]^{-1}.
\end{equation}
Again, for $\delta=0$ we return to Essam's result. In Fig.~1 we plot
Eq.~(4). For all $\delta$ we obtain $q_c=1/2$. But while for $\delta<1$
the transition to percolation is second order, we find a tricritical
point at $\delta=1$ and a first order transition for $\delta>1$. The
discontinuity is equal to $1/\zeta(\delta)$. Such a model can be
interpreted as directed compact percolation in a spherical configuration
where $n$ refers to the radius of the percolating bubble in dimension
$D=\delta+1$. Other choices of $P_{out}$, $P_{in}$, and $P_{stay}$ 
might lead to further interesting interpretations.

As an application of the ideas presented in the previous paragraphs, we
discuss the adsorption fraction for an extended polymer growing near an 
attractive cylindrical boundary.\cite{Bo} It highlights the profound impact
of a curved boundary on the critical behavior of physical quantities 
that a planar approximation could not reveal.

In outlining the theory we extend on the beautiful treatment in
Refs.~2 and ~17. Consider a lattice consisting of an infinite 
set of concentric cylinders of unit spacing. Let the innermost cylinder 
-- the surface of the boundary -- be of integer radius $m\geq 0$, the 
next innermost of radius $m+1$, and so on. Each
cylinder is labeled by its radius. Now consider a random walk in unit
steps either parallel to the length or perpendicular to these cylinders,
starting on the boundary. A parallel step is taken with a relative
weight of $P_{stay}\equiv 1$, steps outward and inward are taken with
relative weights of $P_{out}(n)=2(n+1)^{\delta}
/[n^{\delta}+(n+1)^{\delta}]$ and $P_{in}(n)=2n^{\delta}
/[n^{\delta}+(n+1)^{\delta}]$, respectively, for $\delta=1$ and
$n>m$,\footnote{For $\delta=2$ we can study the case of a spherical
boundary while for $\delta=0$ or $m\to\infty$ we recover the case of a
planar boundary. Negative $\delta$ would correspond to a polymer
growing in a cavity.} while on the boundary $P_{out}(m)=1,~P_{in}
(m)=0$. At each step the walker picks up a statistical
weight $z$, while for each step on the boundary the walker also acquires
an additional weight of $\kappa\geq 1$. For simplicity, we neglect 
self-interaction and excluded-volume effects.

A walk with $L>0$ parallel steps has reached $L$ levels,
$\{h_i\}_{i=1}^{L+1},~h_i\geq m$, above or on the boundary. We want to
restrict these walks such that $\vert h_{i+1}-h_i \vert \leq 1,~0\leq
i\leq L$.\footnote{This restriction on the directed walk will have no impact 
on the scaling behavior for $L\to\infty$.} The transfer
matrix $T_{h_{i+1},h_i}$ that describes the transition of the walker
from the $i$th to the $(i+1)$st level is given by
\begin{equation}
T_{j,i}=z^{\vert j-i\vert} \kappa^{\delta_{m,j}}\left [P_{stay}
\delta_{j,i} + P_{out}(i) \delta_{j-1,i} + P_{in}(i) \delta_{j+1,i}
\right ].
\end{equation}
The partition function $Z_L$ for walks extending $L$ parallel steps is given by
\begin{equation}
Z_L=z^L\sum_{h_i}\delta_{m,h_0} T_{h_1,h_0} T_{h_2,h_1}\ldots T_{h_L,h_{L-1}} 
 =z^L {\vec b}^{(t)} T^L {\vec e}.
\end{equation}
The total partition function for walks of all length,
$Z=\sum_{L=1}^{\infty} Z_L$, then evaluates to
\begin{equation}
Z(z,\kappa)=z {\vec b}^{(t)} T (1-zT)^{-1} {\vec e}.
\end{equation}
If $\lambda_{max}$ is the largest eigenvalue of $T$, then $Z$ diverges
for $z\nearrow z_{\infty}(\kappa)=1/\lambda_{max}$.

The average length of a walk and the average number of steps taken on the 
surface of the boundary are usually defined to be 
\begin{equation}
<N(z,\kappa)>=z \partial_z \ln{Z(z,\kappa)},
\qquad <N_s(z,\kappa)>= \kappa \partial_\kappa \ln{Z(z,\kappa)},
\end{equation}
respectively. Both, $<N>$ and $<N_s>$, diverge 
for $z\nearrow z_{\infty}(\kappa)$, defining an infinitely long walk.
We want to interpret such walks with the infinite chain
limit of polymers stretched out along a cylindrical
boundary. Then, $<N_s>$ refers to the number of monomers which are adsorbed
on the boundary as a function of the attractive potential $\kappa$.
The fraction of adsorbed monomers $P(\kappa)$ is given by
\begin{equation}
P(\kappa)=\lim_{z\nearrow z_{\infty}(\kappa)} {<N_s(z,\kappa)> \over
<N(z,\kappa)>}=-{\kappa \over z_{\infty}(\kappa)}{d z_{\infty}(\kappa)
\over d\kappa}.
\end{equation}
Thus, $z_{\infty}(\kappa)$ marks a line in the $(\kappa,z)$-plane for
which $P(\kappa)$ is defined. To obtain a non-vanishing adsorption
fraction it is necessary that the attractive potential 
$\kappa$ is larger than some critical value, $\kappa_c$. In terms of the 
eigenvalues $\lambda=\lambda(z,\kappa)$ of the transfer matrix $T$, 
$\kappa_c$ is found to be the smallest value of $\kappa$ 
for which $T$ has a bound state on the line $z=z_\infty(\kappa)$.

The spectrum $\lambda$ of the transfer matrix $T$ is determined by
the eigenvalue problem
\begin{equation}
\lambda g_n=\sum_{i=m+1}^{\infty} T_{n,i}\,g_i =\cases{
g_n+z {2n\over 2n-1} g_{n-1}+z {2(n+1)\over 2n+3} g_{n+1},& $n\geq m+2$;\cr
\noalign{\smallskip}
g_{m+1}+z g_m+z {2(m+2)\over 2m+5} g_{m+2},& $n=m+1$;\cr
\noalign{\smallskip}
\kappa g_m+\kappa z {2(m+1)\over 2m+3} g_{m+1},& $n=m$.\cr}
\end{equation}
This system of equations is most conveniently examined with generating
function techniques: Defining $H(x)=\sum_{n=m+1}^{\infty} x^n g_n/(2n+1)$,
we obtain a first order linear, inhomogeneous differential equation for
$H$. Bound states of $T$ are determined by requiring both, that 
$g_n\to 0$ for $n\to\infty$ and that $g_n$ fulfills the boundary 
condition at $n=m$.
This eigenvalue condition is equivalent to requiring that $H(x)$
may have no singularity for $|x|<1$. Writing $\epsilon=2z/(\lambda-1)$ and
$\gamma=(1-\sqrt{1-\epsilon^2})/\epsilon$, we find
\begin{equation}
{z\kappa\epsilon\over \gamma}=\left[(2z+\epsilon)(\kappa-1)-{z\kappa\over m+1}
\right]{m+1\over m+1/2} {F({1\over 2},m+1;m+{3\over
2};\gamma^2)\over F({1\over 2},m;m+{1\over 2};\gamma^2)},
\end{equation}
where $F$ is a hypergeometric function. We obtain $z_{\infty}(\kappa)$
implicitly from Eq.~(11) by replacing $\lambda=1/z_{\infty}(\kappa)$.
If $\lambda\geq 1+2z$, or equivalently $\kappa\geq \kappa^*(z)=
(1+2z)/(1+z)$, there is one solution of Eq.~(11) for
$\lambda=\lambda_{max}$ such that both conditions on the eigenvectors
$g$ are met. We obtain $\kappa_c=4/3,~z_c=1/2$ at the intersection of
$\kappa^*(z)$ and $z_{\infty}(\kappa)$. Inserting
$z_{\infty}(\kappa)=z_c-\Delta z(\kappa)$ and $\kappa=\kappa_c+\Delta\kappa$
into the equation for $z_{\infty}$, we get in the limit $\Delta
z\to 0_+,~\Delta \kappa\to 0_+$:  
\begin{equation}
\Delta z(\kappa)\sim {1\over 48}
e^{-{8\over 9(m+1)\Delta \kappa}},\quad \Delta \kappa\ll 1/(m+1),
\end{equation}
neglecting exponentially smaller corrections. Using Eq.~(9), we arrive
at Eq.~(1). In Fig.~2 we plot the exact values of $P(\kappa)$ for
$m=0,~1,~2,~3$ and $m=\infty$ for $\kappa\geq\kappa_c=4/3$.
Note the substantial error in estimating $\kappa_c$ for small $m$ using
an extrapolation from values where $P(\kappa)$ appears to be linear. A 
measure of the overestimate on $\kappa_c$ is given by the inflection point of
$P(\kappa)$ in Eq.~(1) at $\Delta\kappa= 0.188/(m+1)$, beyond which 
the approximation in Eq.~(1) ceases to be valid and $P(\kappa)$ grows
linearly. 

The simplicity of the dynamics in this new random walk model, and
the nontrivial scaling obtained from it, raise interesting questions
regrading the universal properties of this lattice. At the critical 
transition only a few fundamental properties of the system determine 
its behavior. In this model, the critical behavior arises from the
balance between a short-range attractive potential and the spatial entropy.
We argue that these features are sufficiently well represented by a random
walk on a hyper-spherical lattice.
It has been shown that such a lattice reproduces all the universal scaling 
properties expected of lattices.\cite{BeBoMo} The critical behavior 
obtained on this lattice for rotationally symmetric systems should 
therefore reflect the universal critical behavior of the system. The 
advantage of random walks on hyper-spheres is to describe the critical 
behavior in a minimal and tractable way in comparison, for example, to a 
far more structured hyper-cubic lattice.

We would like to thank many of the participants of the SERIES'94 conference 
at the Technion in Haifa for inspiring some of the applications for our
model. In particular, we thank Vladimir Privman and Tony Guttmann for
useful discussions. One of use (SB) would like to thank the Institute for Theoretical
Physics at the Technion for its
hospitality. MM is supported in part by  GIF, the Winnipeg Research Fund and 
the Fund for Promotion of Research at the Technion. SB is supported by the 
U.S. Department of Energy under Contract No. DE-AC02-76-CH00016.

\newpage
\section*{FIGURE CAPTIONS}

\noindent
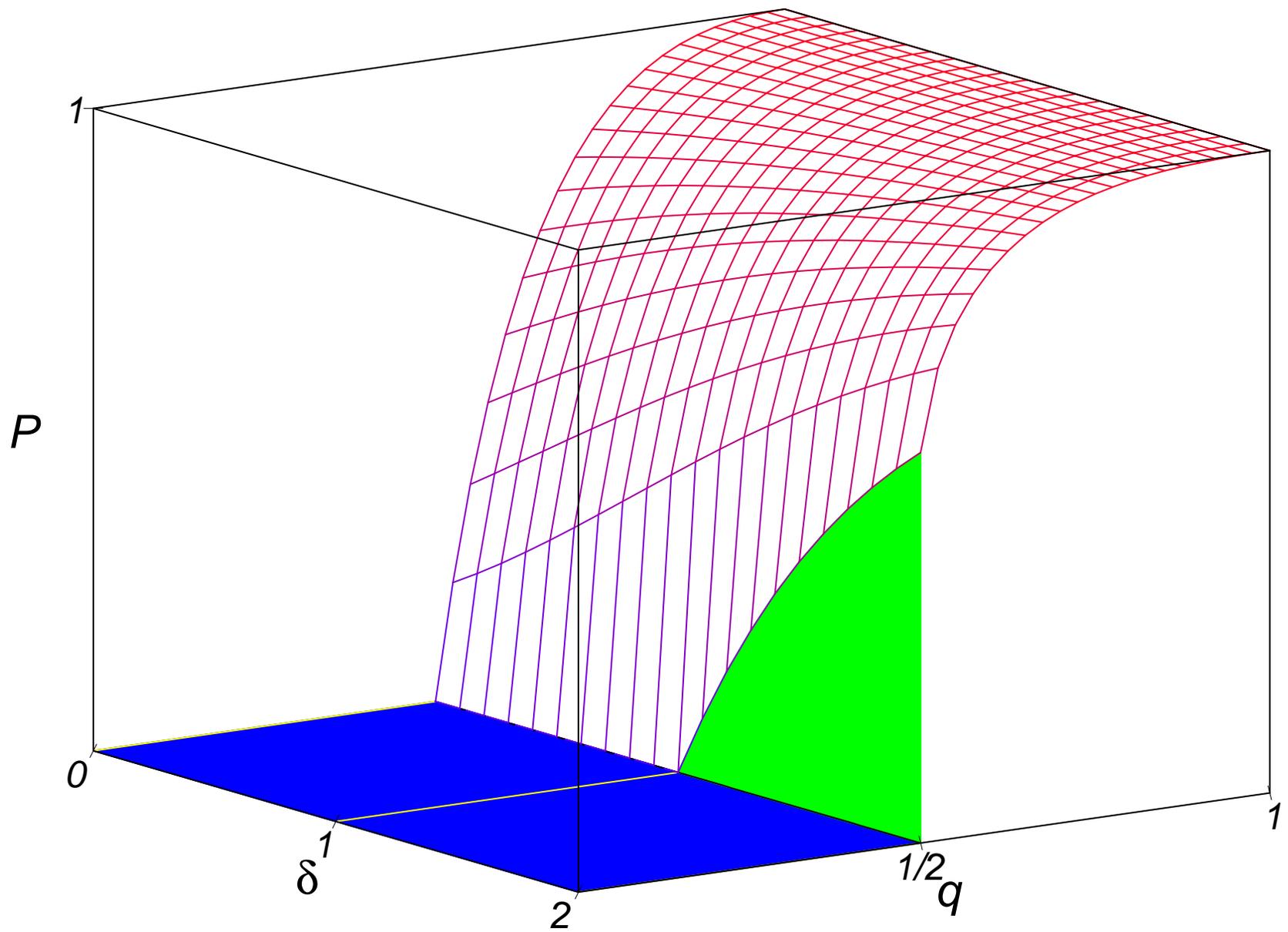
FIGURE 1: The percolation probability $P(q,\delta)$ for directed compact
percolation in a curved geometry. For any $\delta$, the percolation
threshold occurs at $q_c=1/2$, but for $\delta<1$ the transition is
second order, while for $\delta>1$ the transition is first order with a
discontinuity of $P(q_c,\delta>1)=1/\zeta(\delta)$, indicated by a
darkened wedge.
\bigskip

\noindent
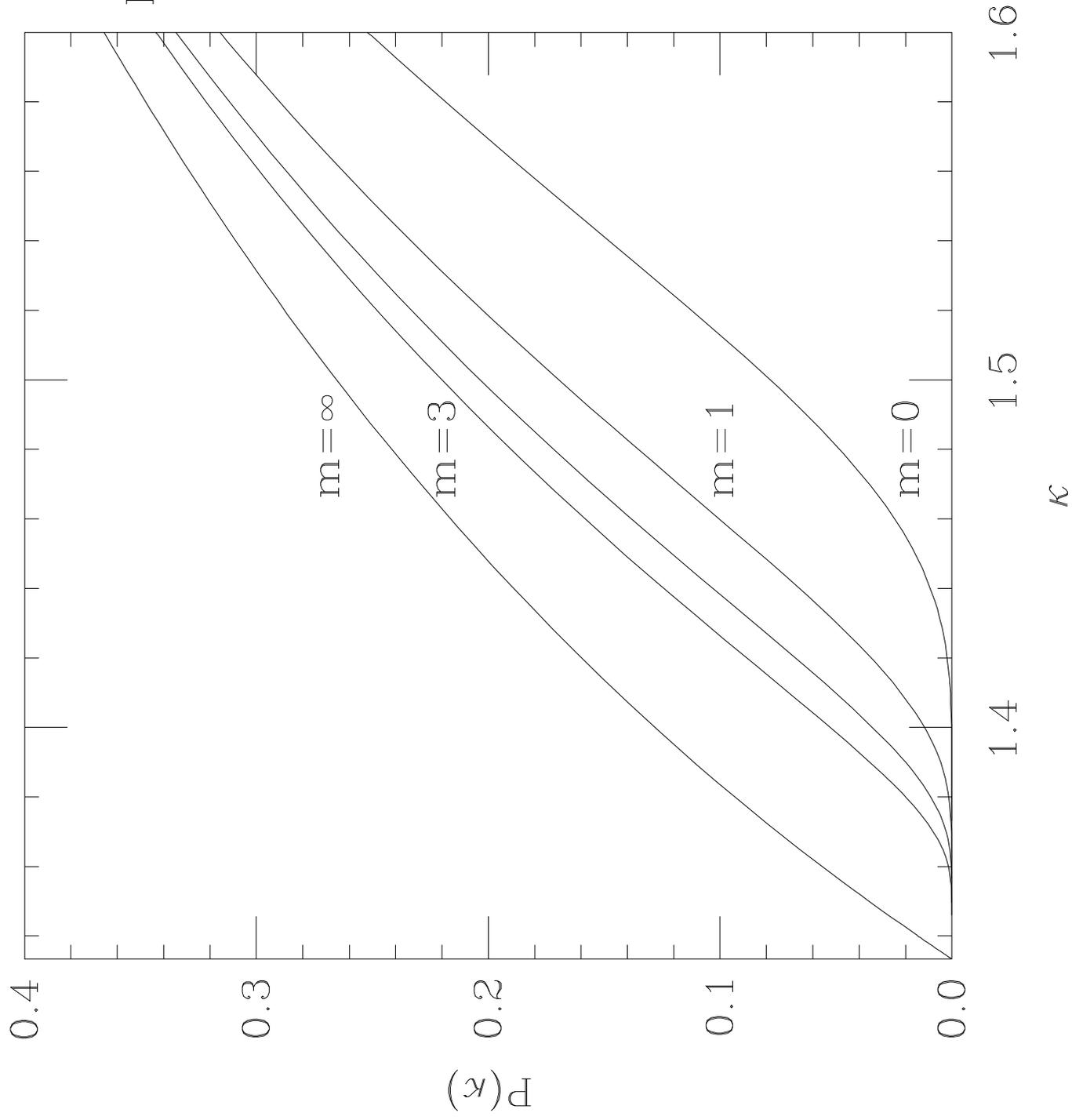
FIGURE 2: The exact adsorption fraction $P(\kappa)$ plotted for 
$m=0,~1,~2,~3$ and $m=\infty$ and $\kappa\geq\kappa_c=4/3$.
{}For finite $m$, $P(\kappa)$ vanishes
exponentially for $\kappa-\kappa_c\to 0_+$ while it crosses over to a linear 
increase when $\kappa-\kappa_c> 0.188/(m+1)$. For $m=\infty$ we recover the
linear scaling for $P(\kappa)$ that was found in Ref.~17.


\newpage

\begin{thebibliography} {99}
\bibitem{AIP} See for example {\it Random Walks and Their Application 
in the Physical and Biological Sciences}, eds. M. F. Shlesinger 
and B. J. West (American Institute of Physics, New York, 1984).
\bibitem{Priv1}
V. Privman and N. M. \v Svraki\'c, {\it Directed Models of Polymers,
Interfaces, and Clusters: Scaling and Finite-Size Properties}, (Springer,
Berlin, 1989).
\bibitem{Aizenman}
M. Aizenman, Phys. Rev. Lett. {\bf 47}, 1 (1981), and Commun. Math.
Phys. {\bf 86}, 1 (1982); and J. Fr\"ohlich, Nucl. Phys. B {\bf 200}
[FS4], 281 (1982).
\bibitem{Sokal}
R. Fernandez, J. Fr\"ohlich, and A. D. Sokal, {\it Random Walks,
Critical Phenomena, and Triviality in Quantum Field Theory},
(Springer, Berlin, 1992).
\bibitem{FloMo} P. J. Flory, J. Chem. Phys. {\bf 17}, 303 (1949);
E. W. Montroll, J. Chem. Phys. {\bf 18}, 734 (1950).
\bibitem{Binder}
K. Binder and K. Kremer, in {\it Scaling Phenomena in Disordered Systems},
eds. R. Rynn and A. Skjeltorp (Plenum, New York, 1985).
\bibitem{Dimarzio1}
E. A. Dimarzio and F. L. McCrackin, J. Chem. Phys. {\bf 43}, 539 (1965).
\bibitem{Taka}
G. H. Weiss and R. J. Rubin, Adv. Chemical Physics {\bf 52}, 363 (1983);
A. Takahashi and M. Kawaguchi, Adv. Polymer Sci. {\bf 46}, 1 (1982);
Yu. S. Lipatov and L. M. Sergeeva, {\it Adsorption of Polymers}, (Halsted,
Jerusalem, 1974).
\bibitem{Dimarzio2}
E. A. Dimarzio and M. Bishop, Biopolymers {\bf 13}, 2331 (1974).
\bibitem{BeBoMe}
C. M. Bender, S. Boettcher, and L. R. Mead, J. Math. Phys. {\bf 35},
368 (1994).
\bibitem{ItDr}
C. Itzykson and J. Drouffe, {\it Statistical Field Theory,} Vol. 1, pp.
1, (Cambridge University, Cambridge, 1989).
\bibitem{BeBoMo}
C. M. Bender, S. Boettcher, and M. Moshe, J. Math. Phys. {\bf 35}, 4941 
(1994).
\bibitem{DoKi}
E. Domany and W. Kinzel, Phys. Rev. Lett. {\bf 53}, 311 (1984).
\bibitem{Essam}
J. W. Essam, J. Phys. A: Math. Gen. {\bf 22} (1989) 4927.
\bibitem{Bateman}
For a discussion of the function $\Phi$, consult the {\it Bateman Manuscript
Project,  Higher Transcendental Functions}, ed. A. Erd\'elyi, Vol.
1, pp. 27, (McGraw-Hill, New York, 1953).
\bibitem{Bo}
For more details, see S. Boettcher, Phys. Rev. E., (to appear).
\bibitem{Priv2}
V. Privman, G. Forgacs, and H. L. Frisch, Phys. Rev. B {\bf 37}, 9897
(1988).

\end{thebibliography}
\end{document}